\documentclass[%
 reprint,
 amsmath,amssymb,
 aps,
prl,
]{revtex4-2}

\usepackage[
  paperwidth=210mm,   
  paperheight=297mm,  
  top=20mm,           
  bottom=20mm,        
  left=17mm,          
  right=15mm,         
  columnsep=5mm      
]{geometry}

\usepackage{physics}
\usepackage{float}
\usepackage{graphicx}
\usepackage{dcolumn}
\usepackage{bm}
\usepackage{xcolor}
\usepackage{booktabs} 

\usepackage{comment}
\usepackage[font={scriptsize}, justification={justified},singlelinecheck=false]{caption}

\usepackage{ragged2e}

\makeatletter
\long\def\@makecaption#1#2{%
  \par
  \begingroup
    \footnotesize
    \setlength{\parindent}{0pt}
    \justifying
    \noindent #1.~#2\par
  \endgroup
}
\makeatother

\usepackage[english]{babel} 
\usepackage{lipsum}
\usepackage{graphicx}
\usepackage{amsmath}

\begin{document}

\title{Continuous-wave laser absorption spectroscopy of the Thorium-229 nucleus}

\author{I. Morawetz$^1$}
\author{T. Riebner$^{1,2}$}
\author{L. Toscani De Col$^1$}
\author{F. Schneider$^1$}
\author{N. Sempelmann$^1$}
\author{F. Schaden$^1$}
\author{M. Bartokos$^1$}
\author{G. A. Kazakov$^1$}
\author{S. Lahs$^1$}
\author{K. Beeks$^1$}
\author{B. Gerstenecker$^1$}
\author{A. Grüneis$^1$}
\author{M. Pimon$^1$}
\author{T. Schumm$^1$}
\email{thorsten.schumm@tuwien.ac.at}
\affiliation{$^1$Vienna Center for Quantum Science and Technology, Atominstitut, TU Wien, Vienna, Austria}
\affiliation{$^2$Bundesamt für Eich- und Vermessungswesen (BEV), Vienna, Austria}

\author{V. Lal$^3$}
\author{G. Zitzer$^3$}
\author{V. Petrov$^{4}$}
\author{J. Tiedau$^3$}
\author{M. V. Okhapkin$^3$}
\author{E. Peik$^3$}
\email{ekkehard.peik@ptb.de}
\affiliation{$^3$Physikalisch-Technische Bundesanstalt (PTB), Braunschweig, Germany}
\affiliation{$^4$Max-Born-Institute for Nonlinear Optics and Ultrafast Spectroscopy, Berlin, Germany}

\date{\today}

\begin{abstract}

A low-energy nuclear transition in the isotope thorium-229 has been excited in thorium-doped crystals with laser light. This opens the perspective towards a highly stable and robust solid-state optical nuclear clock. The required laser radiation at 148 nm wavelength has so far been produced using pulsed laser systems where only a small fraction of the incident photons has been resonant with the narrow nuclear transition. Here we show that the nuclear resonance can be excited with a continuous-wave narrow-bandwidth laser source with a power of less than 1\,nW, and that the resonance signal can be detected in absorption rather than in fluorescence. This eliminates the slow nuclear fluorescence decay from the detection process and offers a considerable advantage for clock operation through fast signal acquisition. The VUV laser source is based on three sequential frequency doublings, starting from a diode laser at 1187\,nm that is well suited for linewidth narrowing and for frequency comparisons with optical atomic clocks.  We use absorption spectroscopy for the quantitative characterization of two different Th-centers in calcium fluoride crystal and measure the isomeric shift between them. One of the centers shows a very small static electric crystal field gradient $<{0.1}$\,V/\AA$^2$, to be compared to gradients in the range of $100$\,V/\AA$^2$ observed earlier. This indicates a center with high symmetry of the ions surrounding the Th nucleus, promising  
nuclear resonance lines that are nearly independent of the lattice spacing.
\end{abstract}


\maketitle

The 8.4\,eV low-energy nuclear transition of Th-229 is investigated for the application as a optical nuclear clock of very high accuracy and stability~\cite{Peik:2003,Beeks:2021,Tiedau:2024, Elwell:2024, Zhang:2024, Elwell2025, hiraki2025laser}. The nuclear resonance may be probed with thorium ions in an ion trap in vacuum or inside a transparent crystal. In solids, Th-229 becomes a test case for laser M\"ossbauer spectroscopy that is sensitive to the interaction between the nucleus and its environment~\cite{Zhang:2024, hiraki2025laser}. In the first experiments on laser excitation of Th-229, the required coherent vacuum-ultraviolet (VUV) radiation at 148\,nm wavelength was generated using four-wave-mixing (FWM)~\cite{Thielking:2023,Tiedau:2024,Elwell:2024,hiraki2025laser} or high-harmonic generation (HHG)~\cite{Zhang:2024} of pulsed laser sources. In these experiments, only a small fraction of the VUV photons were in resonance with the Th nuclei. For the sources based on FWM, the spectral width was several orders of magnitude wider than the nuclear linewidth in the used host crystals~\cite{Tiedau:2024,Elwell:2024,hiraki2025laser}. For the HHG source using a femtosecond frequency comb, the crystal-field-broadened nuclear linewidth in the range of 10-100\,kHz~\cite{ooi2026frequency} was well resolved, but on the order of $10^5$ non-resonant comb modes were present, while only a single mode contributed to the signal~\cite{Zhang:2024, ooi2026frequency}. Laser excitation was detected by observing the nuclear fluorescence, slowly decaying with a time constant of about 600\,s after blocking the radiation impinging on the crystal. 

Exploiting the full potential of a Th-229 solid-state nuclear clock requires a laser source with a linewidth comparable to the crystal-field-broadened linewidth, and a fast, sensitive and robust detection method. These requirements can be fulfilled with the use of a narrow-linewidth continuous-wave (CW) laser, which in contrast to HHG, combines the full laser power within the nuclear transition and therefore enables direct laser absorption spectroscopy with significantly higher signal-to-noise ratio, as demonstrated in this work.

Coherent VUV light can be generated through frequency conversion of laser radiation starting at longer wavelengths. The task is  complex because only a few non-linear optical materials are transparent in the VUV spectral region and possess the properties for phase-matched frequency conversion.
Recently, two different CW laser sources at the nuclear resonance wavelength of Th-229 have been reported: 
An all-solid-state system based on three consecutive steps of second harmonic generation~\cite{Lal:2025} starting from an infrared diode laser at 1187\,nm, and a system based on FWM in cadmium vapor~\cite{Xiao:2026}. 
While the FWM system produced about 100-times higher VUV power, the solid-state system is more compact and requires the frequency stabilization of only one single laser at a technically convenient wavelength of 1187\,nm that can easily be linked with other optical clocks. An essential element of the CW system is the use of a strontium-tetraborate crystal for frequency doubling to 148\,nm~\cite{Lal:2025, Trabs:16}.

\begin{figure*}[ht]
\centering
\includegraphics[width=1\textwidth]{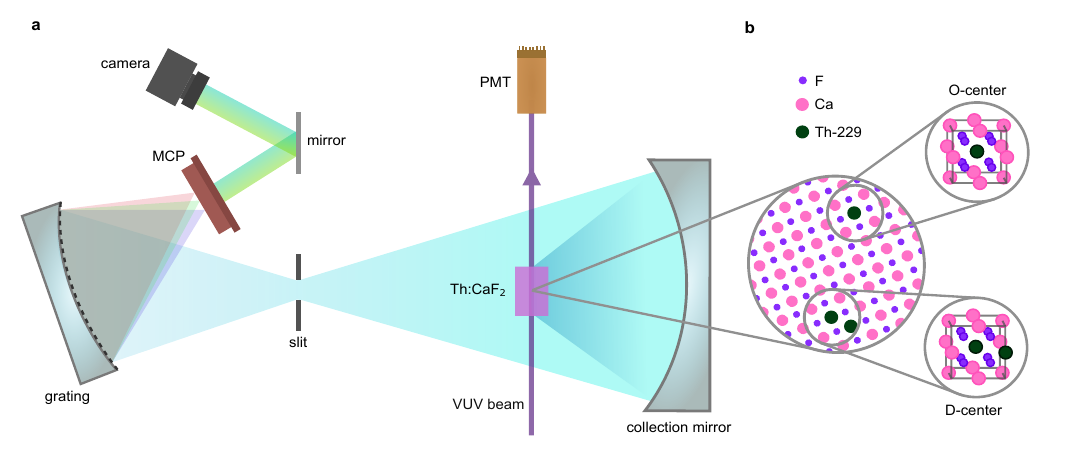}
\caption{\textbf{Illustration of the detection setup and the microscopic structure of investigated Th-229 doping centers in the CaF$_2$ host crystal. a},  For the fluorescence detection, the signal is focused by a collection mirror, diffracted by a spherical concave grating, and detected by a MCP detector with a phosphor screen. The phosphorescence is finally detected using a CMOS camera. For absorption measurements, a CsI PMT is positioned behind the Th:CaF$_2$ crystal. \textbf{b}, Crystal structure of thorium-doped calcium fluoride with the two investigated doping centers. The D-center is a Th dimer configuration with dihedral symmetry and the O-center is a high $O_h$ symmetry configuration. }\label{fig:setup}
\end{figure*}

{\subsection*{Nuclear absorption method}\label{sec2}}

A narrow linewidth CW laser, with a linewidth comparable or narrower than the nuclear transition in the host crystal, has the advantage that all photons can be resonant with the transition. Therefore, the nuclear excitation can be detected in absorption by measuring the attenuation of laser power transmitted through the crystal. A spectroscopic signal can be recorded by modulating the laser frequency over a fraction of the absorption linewidth and detecting the changes in transmitted power in phase correlation with the frequency modulation (see~\cite{bjorklund1983frequency,Kluczynski:2001} for example). The ability to perform absorption detection is pivotal for the operation of a solid-state nuclear clock. 
The absorption detection probes the nuclear excitation, depleting a small fraction of the ground state population, resulting in the attenuation of a directed laser beam while passing the sample.
The measurement is effectively immediate, the timing of the interrogation sequences is determined by optimization of the signal-to-noise ratio, not constrained by the Th-229 excited state lifetime of $\sim\,600$\,s in CaF$_2$ host crystals~\cite{Tiedau:2024,Zhang:2024}. The coherence time for the Th-229 resonance in fluoride crystals is estimated to be below 10\,ms: The coherence time is significantly reduced due to the magnetic dipole interaction of the Th-229 with the surrounding F-19 nuclei~\cite{Rellergert:2010,Kazakov:2012} (see Fig.~\ref{fig:setup}). This sets the maximum excitation time that will lead to the smallest spectroscopic linewidth.

The minimum time for fluorescence detection of the Th-229 nuclei is determined by the desired signal-to-noise ratio and by the time constant of the radiative decay. In preparation of the next interrogation cycle, a waiting time may be required to let the excited stated population return to the ground state, also determined by the radiative decay, potentially accelerated by laser-induced quenching into the range of 100\,s~\cite{Schaden:2025,Terhune:2025,hiraki2025laser} or 10\,s by x-ray-induced quenching~\cite{Hiraki:2024, guan2026x}. This imposes an inefficient operation cycle for the clock where most of the time would be spent for detection and state re-initialization, and only a small fraction of time for interrogation, when the oscillator is actually compared to the nuclear reference and its frequency excursions can be corrected~\cite{Kazakov:2012}. 
Absorption detection improves on the above as the temporal response of the signal is not limited by the long time constant of the radiative decay, and the signal is detected simultaneously with the excitation. 

The use of a narrow-linewidth CW laser and absorption detection offers additional benefits in comparison to the pulsed laser excitation and fluorescence detection used so far. It minimizes the detrimental effects on the signal from non-resonant VUV photons. These effects may comprise AC Stark shift, quenching of the population of the excited state~\cite{Schaden:2025, Terhune:2025, hiraki2025laser} and radiation damage of the crystal~\cite{Beeks:2024}. Any process where resonant nuclear excitation is followed by non-radiative decay (such as shown by~\cite{Elwell2025}) fully contributes to the absorption signal, while remaining undetected in fluorescence. The full spectroscopic information is contained in the amplitude and phase of the transmitted laser beam. Apertures can be used to shield the photodetector from the background of Cherenkov radiation and radioluminescence that is emitted isotropically from the Th-doped crystal. No bulky VUV collection optics are required as for fluorescence detection, where a large solid-angle needs to be covered in order to obtain high sensitivity.
The enhanced sensitivity and rapid response of CW laser spectroscopy allows us to perform high resolution laser M\"ossbauer studies of different Th dopant sites in calcium fluoride (CaF$_2$) and to compare their absorption and excitation spectra. This enables studies on site-specific frequency shift and broadening of nuclear transition lines, temperature dependencies, and effects of external fields and material strain. Laser absorption spectroscopy provides a direct quantitative measurement of the nuclear column density of a specific doping center; it also allows to detect centers which do not decay through VUV radiative decay (or on too short timescales).
\\
\subsection*{Fluorescence and absorption setup}\label{sec2}

The experimental setup consists of a VUV laser source and a nuclear spectroscopy vacuum chamber with thorium doped calcium fluoride (Th:CaF$_2$).

The Th:CaF$_2$ crystal is a segment of the X2 sample~\cite{Beeks:2022}, other pieces of the same ingot were previously used in~\cite{masuda2019x,Hiraki:2024,Tiedau:2024,Zhang:2024,temp-sensi,Schaden:2025,ooi2026frequency,hiraki2025laser,guan2026x}. The crystal has a cylindrical geometry with a 3.1(1)\,mm diameter and 4.2(1)\,mm length and is oriented such that the laser traverses along the centerline. The bulk doping concentration of Th-229, obtained by $\gamma$-spectroscopy and weighing, is determined to be 6.6(5)$\times 10^{15}$\,mm$^{-3}$. A spread of concentration within grown crystals of 10\,\% was observed in radial as well as in axial direction; the local concentration probed by the laser may therefore vary more than the indicated uncertainty. During all experiments reported here, the crystal temperature was kept at 294.7(5)\,K. The optical transmission of the crystal at the nuclear transition wavelength was measured to be 40(5)$\%$ using a VUV monochromator in a broad spectral range~\cite{Beeks:2022,Beeks:2024}.

A high power laser at 1187\,nm is frequency-quadrupled to $\approx$\,500\,mW of 296.8\,nm radiation. This radiation is used in the final single-pass second harmonic generation (SHG) step from 296.8\,nm to 148.4\,nm. This step is based on SHG in a random quasi phase-matched (RQPM)~\cite{baudrier2004rqpm,Trabs:16} strontium tetraborate (SrB$_4$O$_7$, SBO) crystal placed in a vacuum chamber. For stable SHG output power, the SBO crystal is kept under high purity N$_2$ (purity 5.0) environment at 1013\,mbar pressure. The chamber with the SBO crystal is separated from the vacuum beamline by a MgF$_2$ viewport.
The VUV beam is guided to the spectroscopy chamber and aligned to the Th-229 crystal using a photomultiplier tube (PMT) positioned behind the crystal (see Fig.~\ref{fig:setup}). A movable mirror allows to re-direct the beams into a VUV spectrometer (HP Spectroscopy easyLIGHT) for diagnostics and power measurements.
About 1\,nW of VUV power at 148.4\,nm is generated from 350\,mW of fundamental power~\cite{Lal:2025}. It is reduced due to losses from three dichroic mirrors, two MgF$_2$ viewports and a collimation lens. Further losses appear due to Fresnel reflections, significant scattering from both crystal facets, and the VUV transmission of the Th:CaF$_2$ crystal. Therefore, the final VUV optical power measured after the crystal is $\approx$\,70\,pW.

Initially, an overview VUV spectrum of the nuclear quadrupole structure is recorded using the TA-FHG pro laser, frequency stabilized at the wavelength of 1187\,nm to a relevant mode of an infrared frequency comb (Menlo FC1500-250-ULN) by an offset frequency phase lock.
The repetition rate of the frequency comb is stabilized by locking another comb mode to a high-finesse cavity stabilized external-cavity laser (Menlo ORS) at 1542\,nm located at the TU Wien Atominstitut (ATI). The VUV radiation linewidth in this operation mode is in the order of 300\,kHz (see Methods for explanations). The frequency lock of the TA-FHG laser to a comb mode is used for an initial search and a wide scan of the fluorescence and absorption spectra.

Further, for better spectral resolution of quadrupole lines required for the operation of a solid-state
nuclear clock, the TA-FHG pro laser is phase locked to the radiation of a high-finesse cavity stabilized external cavity diode laser (ECDL) at 1187\,nm which has an instability of $\approx$\,10$^{-15}$ at 1\,s. This allows us to  narrow the VUV radiation linewidth significantly. Although we have not directly measured the VUV linewidth, spectral analysis suggests that it lies on the order of 10\,kHz or below.
In both cases the scanning is provided by changing of the phase lock reference frequency.

Measurements of the absolute VUV laser frequency are performed by referencing to a signal from the Austrian Federal Office of Metrology and Surveying (BEV), traceable to either an active H-maser traceable to Coordinated Universal Time (UTC) or an Yb$^+$ single ion clock~\cite{Stuhler26}, depending on availability, delivered to the ATI via a Doppler-compensated fiber link. The optical scheme of the laser system is described in detail in the Methods section.

The detection chamber is designed to capture a large fraction of the fluorescence photons emitted by the Th:CaF$_2$ crystal as well as to monitor the power of the transmitted VUV laser. A schematic is shown in Fig.~\ref{fig:setup}. 
The Th:CaF$_2$ crystal is mounted on a three-axis vacuum translation stage used for precise alignment such that the Cherenkov radiation emitted by the crystal~\cite{Beeks:2022} and the transmitted VUV laser power is maximized. The fluorescence signal is spectrally resolved using a modified high numerical aperture Seya-Namioka spectrometer (HP Spectroscopy) where the emitted light from the Th:CaF$_2$ crystal is collected by an elliptical collection mirror with a diameter of 13.6\,cm, focused onto a 1\,mm slit and then diffracted and refocused by a spherical concave grating. Spatially resolved photon counting is performed by a CsI-coated microchannel plate (MCP) with a phosphor screen on the backside. The phosphorescence is finally detected using a CMOS camera. Additionally, a lead shield is used to protect the MCP from high energy $\gamma$-radiation. 
For absorption measurements, laser alignment, and VUV laser power measurements, a photomultiplier tube (PMT) with a CsI-coated photocathode (Hamamatsu R6835) is mounted behind the Th:CaF$_2$ crystal.

\subsection*{Fluorescence measurements}

\begin{figure*}[ht]
\centering
\includegraphics[width=1\textwidth]
{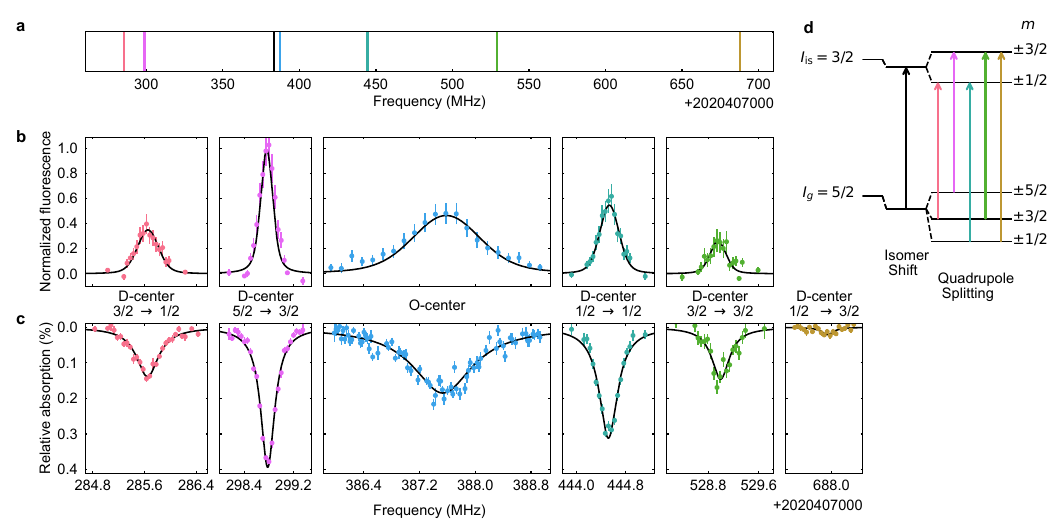}
\caption{\textbf{Fluorescence and absorption spectroscopy of the Th-229 nuclear transition}, showing signals from two Th centers (O and D) and the resolved quadrupole components of the D-center. \textbf{a}, Overview of the detected lines on the MHz scale. The black line shows the position of the calculated center of the quadrupole-split D-center.\textbf{b}, Fluorescence signals (excitation spectra). The 1/2$\rightarrow$3/2 line fluorescence signal is not shown due to low signal-to-noise ratio. \textbf{c}, Absorption signals. 
The O-center line corresponds to a Th defect which has a nearly vanishing electric field gradient at the nucleus position. \textbf{d}, The Th-229 quadrupole structure level diagram of the D-center including the isomer shift. The colored arrows represent the observed transitions in \textbf{a}, \textbf{b} and \textbf{c}.}

\label{fig:measurements}
\end{figure*}
When performing spectroscopy with fluorescence detection, the Th:CaF$_2$ crystal is periodically illuminated for 800\,s, followed by a 800\,s detection interval, dictated by the decay constant of the nuclear excited state. During detection, the laser radiation is blocked by a mechanical shutter to avoid any influence of stray light.
The VUV laser frequency is scanned in 50\,kHz steps over the $\approx$\,1\,GHz range of the quadrupole splitting in Th:CaF$_2$ reported by~\cite{hiraki2025laser,Zhang:2024}. 

We detect 5 lines shown in Fig.~\ref{fig:measurements}b, corresponding to two distinct Th-229 dopant sites in the crystal CaF$_2$ structure (see Fig. ~\ref{fig:setup} and discussion below). Each transition is scanned twice, once from lower to higher frequency and once from higher to lower frequency to account for the asymmetry in the resonance curves due to the long isomer decay time~\cite{Tiedau:2024}. Fig.~\ref{fig:measurements}b shows the corrected line profiles after superposition of data from both scan directions.

\subsection*{Absorption measurements}
In a new approach that is enabled by the narrow-linewidth CW laser, the transmission through the crystal is measured to record the absorption spectrum of the Th-229 nuclei. In contrast to fluorescence, this measurement can be performed continuously because the nuclear ground state population is not depleted significantly and therefore the absorbed VUV laser power is approximately time-independent. The first absorption spectra shown in Fig.~\ref{fig:measurements}c are recorded with the 1187\,nm laser phase locked to the frequency comb mode, and the PMT current is measured with a picoammeter. 
The transmitted power is measured differentially and the absorption is calculated as $(I_0-I)/I_0$ where $I$ is the detected laser intensity on-resonance and $I_0$ is the detected intensity 2\,MHz detuned off-resonance. This differential detection eliminates the influence of slow laser power fluctuations on the signal. A period of 4\,s is used for alternating between the two frequencies.
This measurement cycle is determined by the picoammeter readout. Each data point is averaged over 320\,s. For the data shown in Fig.~\ref{fig:measurements}, the recording time in absorption typically is 5 times shorter than in fluorescence.

\begin{table}[htbp]
\centering
\caption{Fitted spectral line parameters and relative transition strengths. Values of $f_0$ correspond to the central frequencies of the individual transitions, FWHM corresponds to the full-width at half-maximum of the lines and $A$ represents the absorption amplitude of the lines.\\}
\label{tab:line_parameters}
\begin{tabular}{l l l l l}
\hline
Transition & $f_0$ (kHz) & FWHM (kHz) & $A$ (\%)\\
\hline
3/2 $\rightarrow$ 1/2 & 2020407285641(9) & 472(24) & 0.13(1)\\
5/2 $\rightarrow$ 3/2 & 2020407298782(4) & 343(16) & 0.36(3)\\
O-center         
& 2020407387532(14) & 1062(37) & 0.18(1)\\
1/2 $\rightarrow$ 1/2 & 2020407444520(5) & 390(20)  &  0.30(3)\\
3/2 $\rightarrow$ 3/2 & 2020407529003(14) & 382(33)  & 0.14(2)\\
1/2  $\rightarrow$ 3/2 & 2020407687933(38) & 339(73) & 0.02(1)\\
\hline
\end{tabular}
\end{table}

For a simple estimate of the expected photon absorption probability, we calculate {$\rho l \lambda^2 \Gamma_{eg}/(6\pi\Delta\omega)$}~\cite{Wense2020}, where $\rho$ is the Th concentration, $l$ the crystal length, $\lambda$ the nuclear transition wavelength and $\Gamma_{eg}/ \Delta\omega$ is the ratio of the partial vacuum decay rate ($\Gamma_{eg} \approx 2\pi\times 70\, {\rm \mu Hz} \times |C_{I_gm_g1q}^{I_em_e}|^2$, where $C_{I_gm_g1q}^{I_em_e}$ are Clebsch-Gordan coefficients, ranging between $\sqrt{1/15}$ and $\sqrt{2/3}$)  and $\Delta \omega \approx 2\pi \times 400$\,kHz is the broadened linewidth in the crystal. Taking $\rho l\approx 3 \times 10^{22}~{\rm 1/m^2}$, the estimate of the absorption probability 
ranges between 0.15\% and 0.5\% depending on the specific transition. This is in reasonable agreement with the observed relative absorption 
if we take into account the distribution of thorium nuclei over different doping sites. Due to this small cross section only a fraction of $<10^{-7}$ nuclei are in the excited state at any time.

In absorption spectroscopy, we observe six lines due to improved signal-to-noise (Fig.~\ref{fig:measurements}c).
Five lines are associated with a Th dimer defect which has dihedral symmetry~\cite{hiraki2025laser}  (D-center see Fig.~\ref{fig:setup}b) and was first reported in~\cite{Zhang:2024}. The fitted spectral parameters of the lines are presented in Table~\ref{tab:line_parameters}. Some of the measured line centers deviate by more than $3 \sigma$ from previously reported values \cite{Higgins:2024, Fine_structure_Beeks_2025}, which we tentatively attribute to variations of the thorium concentration in the two used segments of the X2 crystal. The observed linewidths vary between 339\,kHz and 472\,kHz and are dominated by the laser linewidth in this set of measurements (see Methods). We find Voigt line profiles to best describe the data in this regime. We fit the electric field gradient (EFG) and nuclear quadrupole moments using our determined line centers with the same method as in~\cite{Fine_structure_Beeks_2025}, and find the best results for $Q_{s}V_{zz} = 335.32(2)\,{\rm eb\, V/}$\r{A}$^2$, $\eta = 0.57183(9)$, $Q_{s}^{m}/Q_{s} = 0.57005(2)$, and $f_{\rm D}=2020407383542(3)\,{\rm kHz}$, where $V_{zz}$ is the largest entry in the diagonalized EFG matrix, by convention defining the $z$ axis, $\eta$ is the asymmetry of the EFG, $Q_{s}^{(m)}$ is the Th-229(m) spectroscopic quadrupole moment, and $f_{\rm D}$ is the frequency of the D-center transition after elimination of the quadrupole splitting.

We additionally observe a line close to the center frequency of the quadrupole structure of the D-center with a larger linewidth of $\approx\,1$\,MHz. This feature has already been reported in~\cite{hiraki2025laser}, with a laser-limited linewidth of 30\,MHz. It can be interpreted as the unresolved quadrupole structure of a Th-center in a high-symmetry doping position in the crystal with nearly vanishing EFG. We use the name O-center because of its $O_h$ symmetry~\cite{Ma:2020}.

\begin{figure}[h]
\centering
\includegraphics[width=\columnwidth]{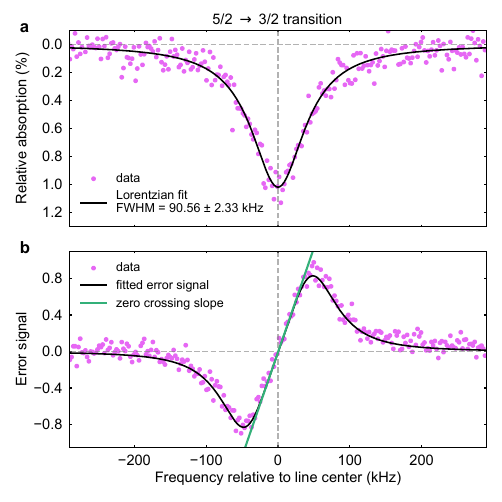}
\caption{\textbf{Absorption and error signals of the D-center 5/2 $\rightarrow$ 3/2 transition observed with the narrow-linewidth} cavity stabilized VUV laser. \textbf{a}, The absorption signal of the  $5/2\rightarrow3/2$ transition. \textbf{b}, The error signal of the absorption line is acquired at a modulation frequency of 10\,Hz with the frequency deviation of 90\,kHz. The signal slope is $2.28\times10^{-5}\Delta\%$/Hz. }\label{fig:ULE}
\end{figure}

The operation of an optical clock requires a signal to stabilize a laser to the reference frequency. This signal can conveniently be produced by 
frequency modulation spectroscopy that is capable of sensitive and rapid measurement of the absorption with narrow spectral features~\cite{bjorklund1983frequency}.
Here, we demonstrate this method on a nuclear transition as a crucial step towards the operation of a nuclear clock.  
To reduce the VUV laser linewidth and noise level, the laser frequency was phase locked to an ECDL at 1187\,nm that is stabilized to a high finesse cavity (see Methods). The PMT readout was switched to photon counting mode, which allowed us to use a 10\,Hz frequency modulation. 
Fig.~\ref{fig:ULE}a shows the absorption profile of the 5/2 $\rightarrow$ 3/2 transition with 3\,s averaging per data point. 
This constitutes an overall reduction of the detection cycle by two orders of magnitude compared to the fluorescence measurement. The spectral line has a full-width at half-maximum of 91(2)\,kHz, which is in agreement with the fluorescence signal width obtained in~\cite{ooi2026frequency} for the X2 crystal. We find spectra in which the laser linewidth is clearly below the nuclear transition linewidth to be best described by Lorentzian profiles.
The detection of the absorption signal in the Th:CaF$_2$ crystal allows us to record an error signal with zero-crossing at the resonance frequency, observed by first harmonic detection (see Fig.~\ref{fig:ULE}b). The error signal is acquired with a frequency deviation of 90\,kHz and averaging the signal over the same integration time as for the absorption profile shown in Fig.~\ref{fig:ULE}a.

\begin{figure}[h]
\centering
\includegraphics[width=\columnwidth]{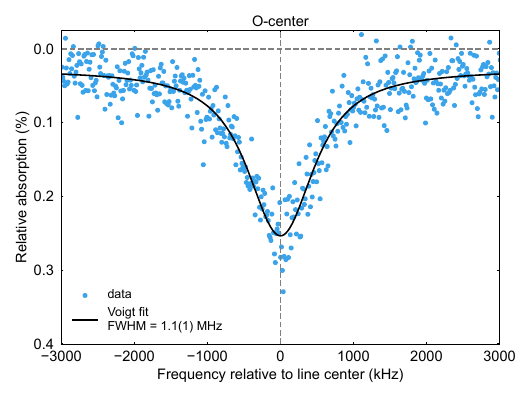}
\caption{ \textbf{Absorption signal of the O-center line.} The absorption signal is recorded with the narrow-linewidth cavity-stabilized VUV laser, similar to D-center line shown in Fig.~\ref{fig:ULE}. No quadrupole structure is observed.}
\label{fig:ULE0EFG}
\end{figure}

The absorption resonance of the O-center recorded with the narrow linewidth laser is shown in Fig.~\ref{fig:ULE0EFG}. For the X2 crystal used in this experiment we do not observe a resolved quadrupole spectrum for this center. From the linewidth and symmetry of the O-center resonance, one can derive an upper bound on the absolute value of the electric field gradient's $V_{zz}$ component of $<0.1$\,V/\AA$^2$ compared to $\approx$\,100\,V/\AA$^2$ for the D-center (see Methods)~\cite{Zhang:2024, hiraki2025laser}. The differences in line broadening observed between the D-center and the O-center will be a subject of further investigations.

\subsection*{Isomeric shift}
The O-center line is offset by $\Delta f_{\rm OD}=f_{\rm O}-f_{\rm D} =3.99(2)$\,MHz from the center of the quadrupole structure of the D-center (black dotted line in Fig.~\ref{fig:measurements}a) determined in the previous section, which we attribute to a difference in the isomer shift~\cite{Gibb:2012}.  
The difference in the isomer shifts indicates that the electron density at the Th nucleus is higher for the O-center than for the D-center. The shift agrees with the results of DFT simulations for the D-center and O-center (see Methods), additionally corroborating the assignment.

\subsection*{Discussion and outlook}

We demonstrate the first precision spectroscopy of the Th-229 nuclear resonance of the O-center and revalidate the proposed microscopic doping structures of the O and D-center by showing agreement between the measured and calculated isomer shift. The O-center line was not reported in laser excitation with a VUV frequency comb~\cite{Zhang:2024, temp-sensi,ooi2026frequency}, but it was observed using a pulsed VUV FWM source~\cite{hiraki2025laser} and in the present experiment. We conjecture that this difference might be related to light induced quenching (LIQ)~\cite{Schaden:2025,guan2026x,Terhune:2025} through non resonant modes in VUV frequency combs.

Further high-resolution spectroscopy of the O-center and other known centers~\cite{hiraki2025laser} can benchmark theoretical models for Th doping in calcium fluoride, further improving on the isomer shift and EFG calculations. Using the O-center for a solid-state nuclear clock should be further investigated with respect to benefits over using the D-center, especially regarding a possibly increased robustness against lattice deformations.
Working with smaller Th doping concentration has led to lower linewidths of $\approx\,30$\,kHz in the case of the D-center, attributed to a lower level of microstrain~\cite{ooi2026frequency}. It will be critical to see if the individual components of the underlying quadrupole structure can be resolved, as for the D-center~\cite{ooi2026frequency}. Measuring the temperature induced frequency shifts and broadening effects will further contribute to the characterization of the O-center.

Most importantly, we demonstrate the first detection of an absorption signal of the Th-229 nuclear transition. Previous experiments based on detection of fluorescence light, were limited by the long decay time of the metastable state. Absorption spectroscopy, in contrast, measures the excitation in real time and is therefore more suitable for stabilizing a laser on the thorium transition. In the future, this could enable the realization of a nuclear clock. Based on the measured linewidth of $\Delta\nu\approx 100\,$kHz for the $(5/2\rightarrow3/2)$ component of the D-center, a detected photon flux $\dot{N}_\gamma\approx 6\times10^{6}\,$s$^{-1}$, 
 and the absorption fraction $A\approx 0.01$, the current setup suggests a fractional frequency instability of $\sigma(\tau)\approx\Delta \nu/(\nu_0 A(\dot{N}_\gamma\tau)^{1/2})\approx 2\times 10^{-12}\sqrt{\tau/\text{s}}$ where $\tau$ is the averaging time and assuming a shot noise limited measurement.

Further improvements in linewidth, detected photon flux, and absorption fraction can be achieved by several means. The interaction length $l$ can be increased by using longer crystals or by integration with an optical cavity. The CW laser power $I_0$ can be increased by using different nonlinear media for the generation of VUV~\cite{Xiao:2026} or cavity enhanced SHG in nonlinear crystals such as SBO or BaMgF$_4$ (BMF). For the present laser source based on SBO~\cite{Lal:2025}, we expect a $\approx$\,10 fold increase in VUV power for cavity-enhanced SHG.  Crystalline host materials different from CaF$_2$, that should show different scaling of nuclear linewidths with concentration, can be explored. A major benefit of a Thorium solid state clock lies in its potential long term stability, using a similar approach as optical frequency standards based on resonances in ions or neutral atoms.  All above-mentioned factors together can result in more than two orders improved shot-noise limited clock performance. Therefore, one can expect to reach an instability of $\leq$10$^{-16}$ at $\sim10^{4}$\,s with further improvement at long time scales. 

In conclusion, in this work we overcome the limitation of the fluorescence detection method caused by the long radiative decay time of the isomeric state of Th-229. Absorption spectroscopy with a continuous-wave VUV laser source provides a clear pathway to operating a solid-state nuclear clock of high stability and accuracy.

\subsection*{Acknowledgments}

We would like to thank Thomas Leder, Martin Menzel, and Andreas Hoppmann for their technical support. We would like to thank Martin Steinel, Burghard Lipphardt, Nils Huntemann and Michael Matus for discussions on the frequency stabilization and optical frequency references. We also thank Dieter Hainz, Monika Veit, and Johannes Sterba from ATI radiation safety for their support in handling radioactive samples. We thank the National Isotope Development Center of DoE and Oak Ridge National Laboratory for providing the Th-229 used in this work. We thank Martin Cizek and Ondrej Cip for providing locking electronics and advice on fiber link stabilization.

Part of this work has been funded by the European Research Council (ERC) under the European Union’s Horizon 2020 research and innovation programme (Grant Agreement No. 856415) and the Austrian Science Fund (FWF) [Grant DOI: 10.55776/F1004, 10.55776/J4834, 10.55776/ PIN9526523]. We acknowledge support from the \"Osterreichische Nationalstiftung für Forschung, Technologie und Entwicklung (AQUnet project), from the Deutsche Forschungsgemeinschaft (DFG) – SFB 1227 - Project-ID 274200144 (Project B04), and from the Max-Planck-RIKEN-PTB-Center for Time, Constants and Fundamental Symmetries. The project 23FUN03 HIOC [Grant DOI: 10.13039/100019599] has received funding from the European Partnership on Metrology, co-financed from the European Union’s Horizon Europe Research and Innovation Program and by the Participating States. The Vienna team acknowledges funding by Defense Advanced Research Projects Agency (DARPA) under grant number HR0011-25-2-0031. 

\newpage
\section*{Methods}\label{meth}

\subsection*{Experimental apparatus and VUV beam guiding}\label{beam}

\begin{figure}[h]
\centering
\includegraphics[width=\columnwidth]{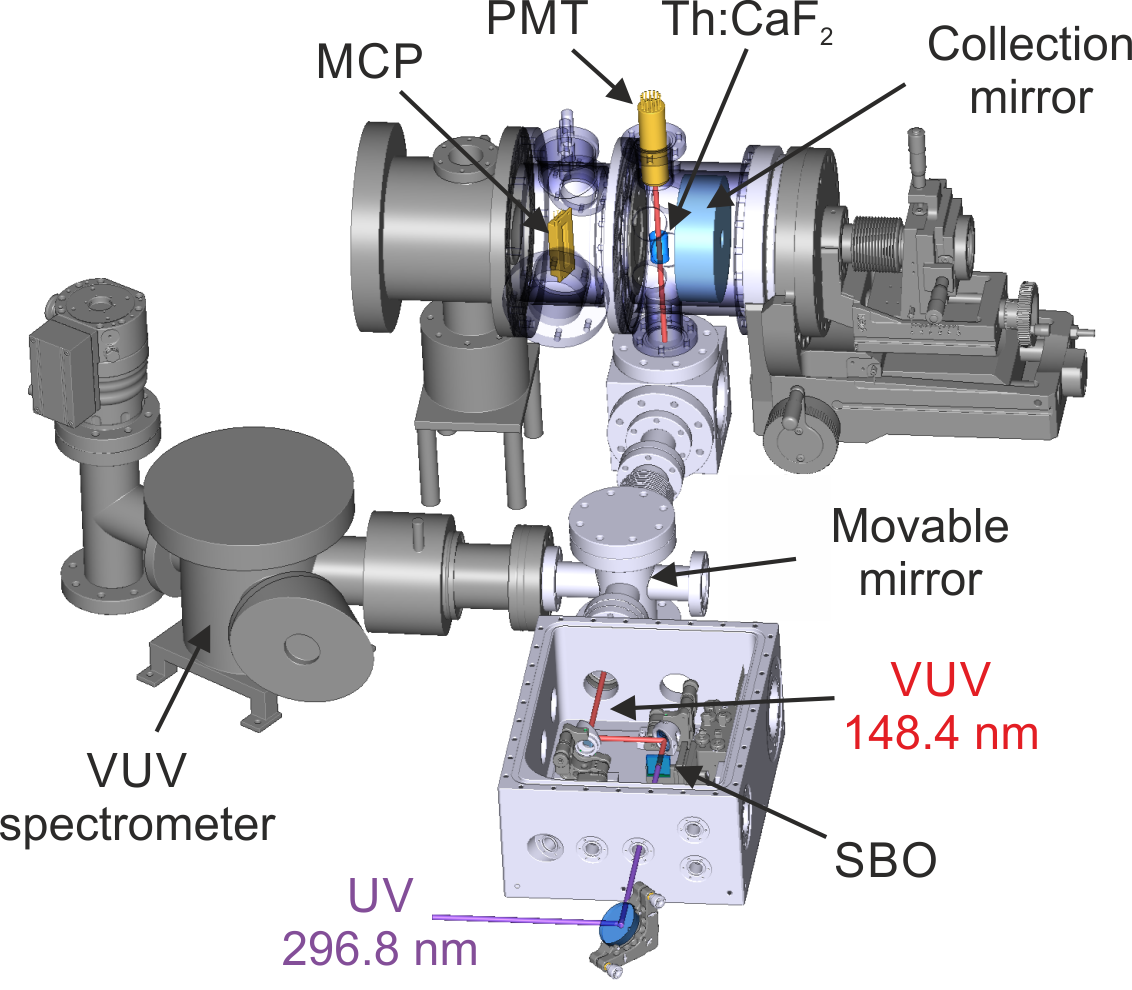}
\caption{\textbf{The experimental apparatus.} The VUV light is generated via a single pass SHG in a RQPM SBO crystal and directed to the Th-doped crystal. The VUV spectrometer is used for optimization and diagnostics of the VUV laser. } \label{fig:setup_cw}
\end{figure}

The design of the spectroscopy apparatus is shown in Fig.~\ref{fig:setup_cw}.
It consists of the chamber with the SBO crystal for SHG frequency conversion, a vacuum beamline, a VUV spectrometer, and a spectroscopy chamber with a Th-doped crystal and the detection system.
The generated VUV beam is aligned to the Th-229
crystal by one plane and two curved dichroic mirrors mounted on motorized mounts using UV radiation as a pilot beam. 
The incidence angle of the radiation on all of the mirrors is 45$^\circ$.
Each dichroic mirror has $\approx90\,\%$ reflectivity at 148.4\,nm and $\approx99\,\%$ transmission for UV at this angle and therefore separate the generated VUV beam from unconverted 296.8\,nm radiation. 
The CsI PMT has a quantum efficiency of $\leq10^{-5}$ in UV. Therefore, the remaining UV background signal registered on the PMT does not exceed  the 5\,\% level of the VUV power transmitted through the crystal.
A CMOS camera is used for the initial alignment of the pilot beam through the crystal in spectroscopy chamber. 
To switch between paths that guide either to the VUV spectrometer for power measurements or the spectroscopy chamber for crystal experiments, the vacuum beamline is equipped with a movable mirror.

\begin{figure*}[htb]
\centering
\includegraphics[width=1\textwidth]{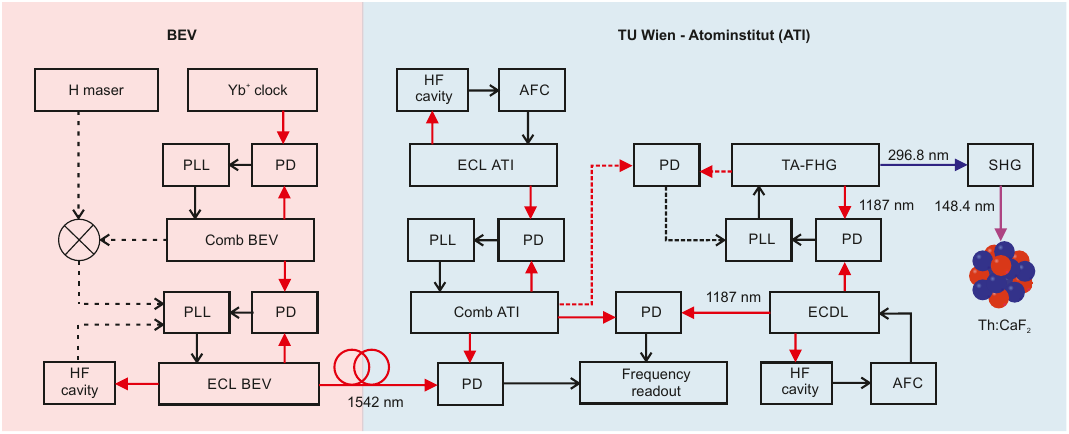}
\caption{\textbf{Optical setup and locking scheme used in our experiment.} Frequency measurements are performed by referencing to a signal from BEV, traceable to either UTC or a Yb$^+$ single ion clock. The TA-FHG laser is phase locked either to a relevant comb mode or to the cavity stabilized ECDL at 1187\,nm.}\label{fig:flasers}
\end{figure*}

\subsection*{Laser frequency stabilization and scanning}\label{meth}

The laser arrangement used for the spectroscopy experiments is shown in Fig.~\ref{fig:flasers}. 
The frequency comb located at BEV based on a Erbium-doped fiber laser is fully stabilized by locking the carrier-envelope offset frequency to a radiofrequency reference and by locking the repetition rate to either an external cavity laser at 1542\,nm (ECL BEV) de-drifted with a feedback loop to an active H-maser traceable to Coordinated Universal Time (UTC) or to a Yb$^+$ single ion clock~\cite{Stuhler26}. In case the Yb$^+$ clock is used as reference, the radiation frequency of ECL BEV is stabilized by locking to a relevant comb mode.  ECL BEV is then guided through a length stabilized optical fiber link to ATI. The second optical comb located at ATI is also fully stabilized by locking to a high finesse cavity stabilized fiber laser and a radiofrequency reference. The long term drift of the comb repetition rate is compensated by analyzing a beat signal with ECL BEV.
Frequency stabilization and scanning of the high power laser system at 296.8\,nm  are provided by a phase lock loop (PLL) of its radiation to a relevant ATI comb mode (dashed lines) or by a phase lock to the radiation of a frequency-stabilized ECDL at 1187\,nm. The ECDL is frequency-stabilized by an automatic frequency control locking system (AFC).
In both cases the scanning is provided by changing of the PLL reference frequency. The frequency chain has a systematic frequency uncertainty of $\pm$1\,kHz in the VUV.

The nuclear transitions full-width at half-maximum (FWHM) spectral width of $\gtrsim$\,300\,kHz observed by locking of the high-power 1187\,nm laser directly to a comb mode is similar to the value reported in~\cite{Zhang:2024}, and is most probably
limited by phase noise of the reference comb mode transferred by the PLL to the upconverted light linewidth.
Therefore, the VUV linewidth in this operation mode is in the order of 300 kHz.

For the high power laser phase locked to the high finesse cavity stabilized ECDL we detected a 91(2)\,kHz spectral FWHM of the $5/2\rightarrow3/2$ transition for the X2 crystal, in agreement with~\cite{ooi2026frequency}. While we did not directly measure the laser linewidth in VUV, we give here only our estimations based on the width of detected spectroscopy signals. The actual laser linewidth is expected to be $\leq$\,10\,kHz  assuming the same transition spectral width observed in earlier experiments using an optical comb~\cite{ooi2026frequency} with $\approx$\,1\,kHz VUV linewidth.

\subsection*{O-center linewidth estimation}

The linewidth of the O-center of 1.1(1)\,MHz (Fig.~\ref{fig:ULE0EFG}) clearly exceeds the laser linewidth in all measurements. We have verified, by performing broadband scans, that it is not a single transition belonging to a quadrupole structure of another Th-defect center in the crystal, analogous to previous work in~\cite{hiraki2025laser}. We conjecture that it is the unresolved quadrupole structure of a Th-center in a high-symmetry dopant site with small or fully vanishing EFG. To constrain the maximum static EFG in terms of $V_{zz}$, we have fitted the O-center line with a set of 6 quadrupole transitions with the respective transition strengths~\cite{Beeks:2025}. 
For modeling, we assume a Cauchy-Lorentz distributed EFG, characterized by a distribution width $\delta V_{zz}$ (half width at half maximum) and a distribution center $V_{zz}$, which describe EFG fluctuations and the static field gradient contribution, respectively. We justify this distribution based on the scaling behavior of the EFG ($V_{zz} \propto r^{-3}$), and by assuming that independently distributed point-defects induce fluctuations in the EFG at the Th site, where the probability of such a point-defect being located in a spherical shell with width $\dd r$ is $p(r) \dd r \propto r^2 \dd r$. Changing $r$ to $V_{zz}$ yields the probability $p(V_{zz}) \dd V_{zz}\propto V_{zz}^{-2} \dd V_{zz}$. The Cauchy-Lorentz distribution provides the correct scaling at its tails and, in addition, is symmetric about the origin. In our fitting procedure, the $\eta$ parameter was fixed to 0.57 (D-center value) and we obtain $\delta V_{zz}={0.928(1)}$\,V/\AA$^2$ and $V_{zz}={0.02(4)}$\,V/\AA$^2$, with the errors being extracted from the relevant elements of the covariance matrix. An upper bound of $\delta V_{zz} = 0.1$\,V/\AA$^2$ also holds when fixing $\eta$ to either end $0$ or $1$.
The very low $V_{zz}$ value indicates a nucleus in a defect center with $O_h$ symmetry. The fluctuations in $\delta V_{zz}$ exceed those found for the D-center in a similar analysis by a factor $\approx\,10$. We note that the fluctuations introduced by $\delta V_{zz}$ constitute a generic model for inhomogeneous broadening, allowing us to compare different defect centers, while the underlying physical mechanisms governing the observed linewidths of the D- and O-centers (and their observed concentration dependence) remain to be identified. 

\subsection*{Calculations of isomer shifts using density functional theory}

\begin{figure*}[t!]
    \centering
    \includegraphics[width=1\textwidth]{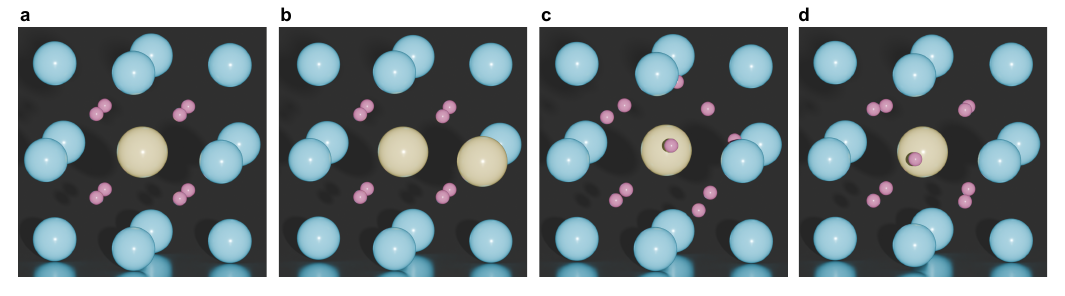}
    
    \caption{\textbf{Relaxed structures of the investigated defects in this study.} The calcium atoms are represented in blue, fluorine atoms in pink, and thorium atoms in yellow. From left to right, the panels show: \textbf{a}, the O-center ($O_h$ defect point group symmetry), \textbf{b}, the D-center ($D_{2h}$), \textbf{c}, two added F ions ($C_{3v}$), and \textbf{d}, a single F interstitial ($C_{4v}$).}
    \label{fig:geometry}
\end{figure*}

To numerically simulate the defect center properties using density functional theory (DFT), we employ a two-step procedure: First, we construct defect centers in a 2$\times$2$\times$2 supercell of the conventional CaF$_2$ unit cell, by replacing one or two (adjacent) Ca ions by Th ions, respectively. We then relax these initial ionic positions to minimize the system's energy. At several steps along the relaxation trajectory, we repeatedly optimize the supercell lattice vectors. After the relaxation criteria are met, we verify that the relaxed structures do not contain imaginary phonon frequencies.

We performed the calculations of the first step using the plane-wave Vienna Ab-initio Simulation Package (VASP)~\cite{vasp1,vasp2,vasp3,vasp4,vasp5} with an energy cutoff of 800\,eV at the $\Gamma$-point. Our convergence criterion of forces imposed a maximum absolute value of 0.00001\,eV/\r{A} on the largest ionic force, while we halted volume optimizations when the energy difference between subsequent iterations was less than $10^{-7}$\,eV. We used the phonopy software~\cite {phonopy1,phonopy2} to compute phonon band structures. The resulting calculations revealed that neither structure exhibited states with imaginary frequencies, aside from inherent numerical inaccuracies. As a result, we concluded that our simulations had converged to the global structural minimum.

In the second step, we calculated the isomer shift for thorium using the electron density difference at the smallest grid point within the linearized-augmented-plane-wave basis in the WIEN2k code~\cite{WIEN2k}. For these calculations, we used the \texttt{-prec 2n} setting, and our convergence criterion was a change in the electronic charge density of less than 0.00001 Rydberg atomic units, as specified by the \texttt{-cc 0.00001} option. We performed convergence tests to the \texttt{-prec 1n} setting and found changes in the isomer shift of about $1\,\text{MHz}$. We used the PBE approximation~\cite{GeneralizedGraPerdew1996} for the exchange-correlation potential throughout all calculations.

The isomeric shift between two electronic environments $A$ and $B$ is~\cite{Mossbauer_Spect_Greenw_1971}
\begin{equation}
    \Delta E_{AB} = \frac{e Z}{6 \varepsilon_0} \Delta \rho_{AB} \Delta \langle R^2 \rangle,
\end{equation}
where $\varepsilon_0$ is the vacuum permeability, $e$ is the elementary charge, $Z = 90$ is the nuclear charge, and $\Delta \langle R^2 \rangle = \langle R_m^2 \rangle - \langle R_g^2 \rangle = 0.0107(9)\,\text{fm}^2$ for Th-229, being the average of three reference values~\cite{Fine_structure_Beeks_2025, Laser_spectrosc_Thielk_2018, Laser_spectrosc_Yamagu_2024}. The electronic environment $A$ corresponds to the single Th high-symmetry O-center, $B$ is the dimer-like D-center.

From our simulations, we obtained values of the electronic charge density difference between the O-center and D-center $\Delta \rho_{OD} = 2.436829687 \times 10^{-5}\,e/r_{\text{Th}}^3 - 2.436829589 \times 10^{-5}
\,e/r_{\text{Th}}^3 =  9.755 \times 10^{-13}\,e/r_{\text{Th}}^3$, where $r_{\text{Th}} = 5.7557$\,fm denotes the radius of the Th-229 ground state~\cite{Angeli2013}. We calculated the corresponding energy shift $\Delta E_{OD} = 88.0698614\,\text{THz} - 88.0698579 \,\text{THz} = +3.60(29)\,\text{MHz}$. We determine the experimental reference by subtracting the field-free frequency of the D-center from the central line frequency $f_0$ of the O-center as $+3.99(2)$\,MHz.

We propose that the coexistence of isolated thorium and thorium dimers reflects the stochastic distribution of dopants during crystal growth. At higher doping concentrations dimer formation becomes more probable as the average inter-dopant spacing decreases.

Before the assignment of the observed O- and D-center based on their respective electric field gradients performed in~\cite{hiraki2025laser}, thorium dopant geometries involving local charge compensation through interstitial F ions were discussed~\cite{dessovic2014229thorium,CharacterizatioTakato2025} (see fig.~\ref{fig:geometry} for all investigated defect structures). Initial placement of two interstitial fluorine atoms relaxes the lattice into a C$_{3v}$ point group symmetry around the defect~\cite{CharacterizatioTakato2025}. A defect center involving a single interstitial fluorine and singly charged by removing an electron (to preserve closed shells) remains in a C$_{4v}$ symmetry. Although the calculated EFGs for these centers (respectively, $V_{zz} = -68$\,V/\r{A}$^2$, $V_{zz} = -279$\,V/\r{A}$^2$ and $\eta = 0$ for both) do not match the observed values, we report their isomer shifts for comparison purposes. The simulations yield an isomer shift of $\Delta E_{C_{3v}D} =  0.076(1)\,\text{MHz}$, while $\Delta E_{C_{4v}D} = -32(3)\,\text{MHz}$.

\bibliography{sn-bibliography}

\end{document}